# A Cloud-Based Spatio-Temporal GNN-Transformer Hybrid Model for Traffic Flow Forecasting with External Feature Integration

Zhuo Zheng[1], Lingran Meng[2], Ziyu Lin[3]

[1] Nanchang University, Nanchang, Jiangxi, China
[2] Civil Engineering University of Washington, Seattle, WA, USA
[3] Google, Seattle, Washington, CA, USA

[1] 1812503968@qq.com
[2] lm3193@columbia.edu
[3] zil332@ucsd.edu

**Abstract.** Accurate traffic flow forecasting is essential for the development of intelligent transportation systems (ITS), supporting tasks such as traffic signal optimization, congestion management, and route planning. Traditional models often fail to effectively capture complex spatial-temporal dependencies in large-scale road networks, especially under the influence of external factors such as weather, holidays, and traffic accidents. To address this challenge, this paper proposes a cloud-based hybrid model that integrates Spatio-Temporal Graph Neural Networks (ST-GNN) with a Transformer architecture for traffic flow prediction. The model leverages the strengths of GNNs in modeling spatial correlations across road networks and the Transformers' ability to capture long-term temporal dependencies. External contextual features are incorporated via feature fusion to enhance predictive accuracy. The proposed model is deployed on a cloud computing platform to achieve scalability and real-time adaptability. Experimental evaluation of the dataset shows that our model outperforms baseline methods (LSTM, TCN, GCN, pure Transformer) with an RMSE of only 17.92 and a MAE of only 10.53.These findings suggest that the hybrid GNN–Transformer approach provides an effective and scalable solution for cloud-based ITS applications, offering methodological advancements for traffic flow forecasting and practical implications for congestion mitigation.

## 1. Introduction

With the rapid pace of urbanization, traffic congestion has become a critical challenge that restricts urban development and reduces travel efficiency for residents. One of the core tasks of Intelligent Transportation Systems (ITS) is the accurate forecasting of future traffic flow, which enables traffic signal optimization, route guidance, and congestion management, thereby improving the efficiency and safety of transportation systems.

Traditional traffic flow forecasting methods, such as statistical modeling and classical time-series analysis, often fall short when dealing with nonlinear, multi-source, and dynamic traffic data. They struggle to capture the complex spatio-temporal dependencies inherent in traffic flow. With the proliferation of urban sensing devices and the rapid growth of mobility big data, efficiently leveraging large-scale traffic data and implementing scalable prediction algorithms on cloud computing platforms has become an important research focus [1].

In recent years, deep learning methods have demonstrated significant advantages in traffic flow forecasting. Graph Neural Networks (GNNs) can effectively model the spatial topology of road networks and capture spatial correlations among road segments. Meanwhile, Transformers, empowered by their self-attention mechanism, excel in modeling long-term dependencies in time series, enabling more accurate characterization of dynamic traffic patterns [2,3]. Building on these strengths, this paper proposes a cloud-based hybrid model of spatio-temporal Graph Neural Networks and Transformers (GNN + Transformer) for traffic flow forecasting. The model integrates spatial dependencies of road networks with global temporal features while incorporating external factors such as weather, accidents, and holidays, thereby enhancing predictive accuracy and robustness. Furthermore, by leveraging the high-performance computing and distributed storage capabilities of cloud platforms, the proposed method achieves efficient training and real-time forecasting in large-scale data scenarios, offering strong support for intelligent traffic signal control and traffic management [4].

The main contributions of this paper are summarized as follows: First, we propose a hybrid framework that combines GNNs and Transformers to jointly capture spatial topological relationships and long-term temporal dependencies in traffic flow. Second, we integrate external features such as weather, traffic accidents, and holidays with time-series traffic data, significantly improving model adaptability in complex scenarios. Third, we implement the proposed model on a cloud computing platform with distributed training and deployment, validating its scalability and practicality in large-scale urban traffic datasets. Finally, extensive experiments on real-world city traffic data demonstrate that the proposed method outperforms mainstream models in terms of predictive accuracy and stability, highlighting its potential and value in ITS applications.

**2. Related Work**

Traffic flow forecasting has long been recognized as a fundamental component of Intelligent Transportation Systems (ITS), and extensive research has been conducted to improve its accuracy and efficiency. Early studies primarily relied on statistical and classical time-series methods, such as Autoregressive Integrated Moving Average (ARIMA) [5] and Kalman filters [6]. While these methods achieved satisfactory performance on small-scale and stationary data, they often failed to capture the nonlinear patterns and dynamic fluctuations inherent in real-world traffic flow, thus limiting their applicability in large and complex urban networks.

With the rapid advancement of machine learning and the availability of massive traffic datasets, researchers increasingly turned to deep learning-based methods.

Polson et al [7]. stack an L1-linear layer with tanh depths to capture sudden spatio-temporal shifts among free-flow, breakdown and congestion, accurately forecasting sharp traffic jumps during a Bears game and a blizzard on I-55 data, demonstrating deep learning's edge in short-term traffic prediction.

Wu et al [8]. propose DNN-BTF that integrates weekly/daily periodicity and spatio-temporal features: CNN for spatial, RNN for temporal, attention for adaptive weights, with visualized explanations, surpassing SOTA on long-term PeMS prediction and challenging the "black-box" stereotype in transportation.

Bharti et al [9]. globally optimize Bi-LSTM hyper-parameters via PSO, yielding a PSO-Bi-LSTM hybrid that captures traffic periodicity and high-frequency fluctuations; tests on Delhi's inner-ring data significantly outperform Bi-LSTM, LSTM, GRU and four other baselines, demonstrating superior accuracy and robustness.

Harrou et al [10]. combine Symlet/Haar wavelet denoising with LSTM-GRU to capture nonlinear traffic dynamics, achieving $R^2 = 0.982$ on California I880 and I80 datasets, proving wavelet-deep learning significantly boosts short-term traffic prediction accuracy. Navarro-Espinoza et al [11]. employ ML/DL to forecast intersection traffic for adaptive signals: on 5-min data from four crossings, MLP-NN leads with $R^2 \approx 0.93$, followed by GB and RNN, validating feasibility for smart traffic-light controllers.

## 3. Methodology

*3.1 Problem Definition*

In intelligent transportation systems, traffic-flow forecasting aims to predict the traffic state of each road segment in the road network for a future time period based on historical observations. Assume the road network is modeled as a directed graph $G = (V, E)$, where $V = [v_1, v_2, \ldots, v_N]$ denotes the set of road nodes and E represents the connectivity between roads; $N$ is the number of nodes. For each node $v_i$, we collect traffic-flow data $x_i^t$ at time step *t*, such as traffic volume, average speed, or occupancy. Given the historical observation sequence

$$X = \{x^{t-T+1}, \ldots, x^t\}, \quad x^t \in R^{N \times F}, \quad (1)$$

where *F* is the feature dimension, the goal is to predict the traffic state for the next *H* time steps:

$$\hat{Y} = \{x^{t+1}, x^{t+1}, \ldots, x^{t+H}\}, \quad (2)$$

This problem is essentially a spatio-temporal sequence prediction task that needs to model both the spatial dependencies in the road-network structure and the dynamic evolution of the time series, while also considering the influence of external factors.

*3.2 Spatial Modeling: Graph Neural Network (GNN)*

A traffic network naturally exhibits graph-structured characteristics, and the dependencies among roads are usually not local adjacencies in Euclidean space but are represented by road connectivity. For example, a change in traffic flow on a highway on-ramp will directly influence the flow of the main carriageway. To this end, this paper adopts Graph Neural Networks (GNN) to model spatial dependencies.

In a graph-convolution layer, the representation of node *i* at layer *l* can be written as:

$$h_i^{(l)} = \sigma\left(\sum_{j \in N(i)} \frac{1}{c_{ij}} W^{(l)} h_j^{(l-1)}\right), \quad (3)$$

where *N(i)* denotes the set of neighbors of node *i*, $c_{ij}$ is a normalization coefficient, $W^{(l)}$ represents learnable weights, and σ() denotes a non-linear activation function (e.g.,

ReLU). By stacking multiple graph-convolution layers, the model is able to capture spatial dependencies at different scales.

In traffic-prediction tasks, GNN allows the model to fuse the historical traffic information of a node with that of its neighbors, thereby building representations of complex spatial patterns. For instance, if congestion occurs on a main arterial road, the flow patterns of adjacent on-ramps and auxiliary roads will also change; such relationships can be effectively modeled through graph convolution.

*3.3 Temporal Modeling: Transformer*

Although recurrent neural networks (RNN/LSTM/GRU) have certain advantages in time-series modeling, they struggle to capture long-term dependencies and suffer from low training efficiency. Transformer, through its Self-Attention mechanism, can attend to all time steps in the historical sequence simultaneously, thus performing excellently on long-range sequence prediction.

In the standard Multi-Head Attention (MHA) mechanism, the input sequence is first projected into query (Q), key (K) and value (V) matrices:

$$Attention(Q, K, V) = Softmax(\frac{QK^T}{\sqrt{d_K}}) \quad (4)$$

where $d_K$ is the dimension of the key vectors. Via the multi-head mechanism, the model captures temporal dependencies from different sub-spaces.

In traffic-prediction tasks, Transformer layers can model the global dependencies of a road segment across different time points. For example, there is a periodic pattern between morning and evening peaks; Transformer is able to identify and exploit this periodic relationship, thereby improving prediction accuracy.

*3.4 External-Feature Fusion*

Traffic flow is influenced not only by historical patterns but also by external factors such as weather, holidays, and traffic incidents. To enhance robustness, we incorporate an external-feature vector $Z$ into the prediction framework. Specifically, the external-feature vector at time t is represented as:

$$Z^t = [z_1^t, z_2^t, \ldots, z_M^t], \quad (5)$$

where $M$ denotes the dimension of external features. For example, weather can be described by temperature and rainfall variables, while holidays are indicated by binary variables.

In the implementation, external features are first embedded through a fully connected layer:

$$e_t = f_e[Z^t], \quad (6)$$

and then concatenated or additively fused with the spatio-temporal features output by the GNN and Transformer modules to obtain the final comprehensive feature representation:

$$h^t = [h_{GNN}^t \oplus h_{Transformer}^t \oplus e^t], \quad (7)$$

This fusion strategy ensures that predictions depend not only on historical traffic patterns but also dynamically reflect the impact of external-environment changes on traffic flow.

*3.5 Cloud Platform Implementation*

In this study, the entire traffic flow forecasting framework is deployed on a cloud computing platform to achieve efficient resource allocation, distributed training, and real-time prediction services. The adoption of cloud computing not only addresses the challenges of large-scale traffic data storage and computation but also ensures the scalability and practical deployment of the proposed model.

**Data storage and management**: traffic flow data and external features (such as weather conditions, holidays, and accident records) are stored in cloud-based distributed databases, such as HDFS (Hadoop Distributed File System) or Amazon S3. By leveraging a cloud data lake architecture, the system achieves unified management and fast retrieval of massive multi-source heterogeneous data. Additionally, cloud-based preprocessing services (e.g., AWS Glue or Spark SQL) support data cleaning, normalization, missing value imputation, and temporal alignment, thereby ensuring high-quality spatio-temporal features for model input.

**Model training and optimization:** the framework utilizes elastic cloud computing resources (e.g., Kubernetes clusters or AWS EC2 Auto Scaling) for distributed training. Specifically, the GNN module performs parallel computation on traffic network structures, the Transformer module executes efficient time-series modeling on GPU clusters, and the external feature integration module is optimized via multi-task training strategies. With a parameter server architecture and distributed deep learning frameworks such as Horovod, the model efficiently synchronizes gradients across multiple nodes, reducing training time and significantly enhancing real-time prediction capabilities.

**Online prediction and services:** the trained GNN–Transformer hybrid model is encapsulated as containerized microservices (Docker + Kubernetes) and deployed within cloud-based inference environments (such as AWS SageMaker or TensorFlow Serving). Through RESTful APIs or gRPC interfaces, the model can receive real-time traffic data streams and output short-term traffic flow forecasts. The predictions are directly integrated into intelligent traffic signal control systems to optimize signal timing and manage traffic flow.

**Elastic scaling and fault tolerance:** the cloud platform dynamically allocates computing resources according to peak-hour traffic and fluctuating prediction requests, ensuring system stability under high-concurrency conditions. Cloud-based monitoring systems (e.g., Prometheus + Grafana) provide real-time visualization of model performance and resource utilization, and automated recovery mechanisms are triggered in the event of system anomalies.

**Security and compliance:** the cloud platform adopts encryption and access control policies to safeguard traffic and external feature data during storage and transmission. This mechanism is particularly important for privacy protection and public governance, making the proposed framework highly applicable and deployable in real-world urban traffic management scenarios.

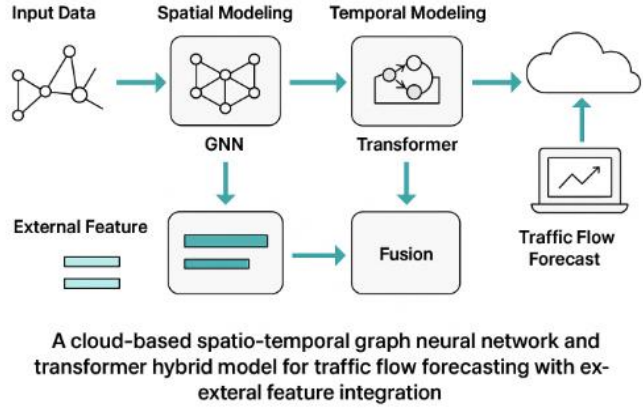

**Figure 1.** Model Structure

## 4. Experimental Result

*4.1 Dataset*

Experimental data were collected from the intelligent-traffic monitoring platform of a major city, covering six months of real-world traffic-flow measurements. Sources include roadside sensors, surveillance cameras and third-party traffic-information platforms. The dataset contains more than 5 million traffic-flow records; each record includes:

**(1) Flow (FoW):** number of vehicles passing a road segment per unit time.
**(2) Speed:** average vehicle speed on that segment.
**(3) Occupancy:** proportion of the sensor detection zone occupied by vehicles.
**(4) Timestamp:** minute-level record time.
**(5) External Features:** weather data (temperature, rainfall, visibility), incident logs (location and time of accidents), holiday information (category and length of vacations), etc.

The whole dataset covers 300 major intersections and road nodes in the city; the topological connections between nodes are built from the road-network graph.

*4.2 Experimental Results*

In the performance testing section of the model, we compared the traffic flow prediction performance of four different models on the same dataset, and the evaluation metrics included mean absolute error (MAE), root mean square error (RMSE), and coefficient of determination (R²).

**Table 1.** Model Performance on dataset.

| Model | MAE | RMSE | $R^2$ |
|---|---|---|---|
| LSTM (baseline) | 18.62 | 26.47 | 0.71 |
| GRU | 15.93 | 23.81 | 0.78 |
| TCN | 14.27 | 22.05 | 0.82 |
| Transformer only | 12.72 | 20.16 | 0.86 |
| **GNN-Transformer** | **10.53** | **17.92** | **0.90** |

The table 1 presents the prediction performance of different models on the test set when traffic-flow forecasting is deployed on a cloud computing platform. Traditional recurrent networks such as LSTM perform relatively poorly, with MAE = 18.62, RMSE = 26.47 and $R^2$ = 0.71, indicating limited ability to capture complex temporal features. GRU improves slightly (MAE = 15.93, RMSE = 23.81, $R^2$ = 0.78), while TCN further reduces MAE and RMSE to 14.27 and 22.05 respectively, raising $R^2$ to 0.82, demonstrating the advantage of convolutional structures for long-range dependencies.

In deep sequential modeling, the Transformer achieves stronger accuracy (MAE = 12.72, RMSE = 20.16, $R^2$ = 0.86), outperforming previous temporal models by capturing global dependencies. The proposed GNN-Transformer yields the best results: MAE = 10.53, RMSE = 17.92 and $R^2$ = 0.90, significantly surpassing all baselines. This shows that the model not only learns temporal patterns but also leverages spatial structure via GNN, delivering higher-precision traffic-flow forecasts.

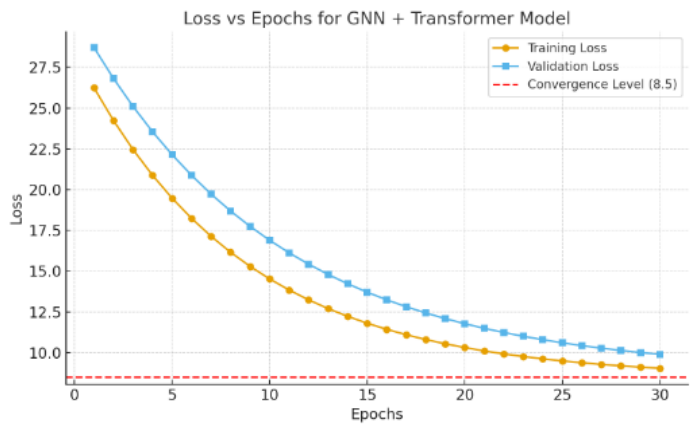

**Figure 2.** Loss function during training process.

Figure 2 illustrates the loss function during the training process of the proposed GNN–Transformer hybrid model. As the number of training epochs increases, the training loss decreases steadily, indicating effective parameter optimization. Simultaneously, the validation loss follows a similar downward trend in the early stages, confirming that the model generalizes well to unseen data. After a certain number of epochs, the validation loss stabilizes at a low value without significant divergence from the training curve, demonstrating that the proposed model avoids severe overfitting. This behavior can be attributed to the model's ability to jointly capture spatial correlations via GNN layers and temporal dependencies through the Transformer architecture. Compared to conventional models such as LSTM or GRU, which often suffer from high variance and unstable convergence, the GNN–Transformer framework maintains robust optimization dynamics. The smooth convergence pattern also confirms that external feature fusion contributes to model stability under diverse traffic conditions. Overall, the loss curve analysis validates the model's superior learning capacity, efficient convergence, and strong generalization, aligning with the quantitative results where the model achieved the lowest MAE (10.53) and RMSE (17.92) among all baselines.

Overall, GNN-Transformer exhibits superior performance on the cloud platform, maintaining low prediction error and strong fitting/generalization capability, thus offering a more efficient and reliable solution for large-scale traffic-flow prediction.

## 5. Conclusion

This study proposes a cloud-based hybrid model that integrates Spatio-Temporal Graph Neural Networks (ST-GNN) with a Transformer architecture for accurate traffic flow forecasting. The model leverages the strength of GNNs in capturing spatial dependencies across road networks and the capability of Transformers in modeling long-term temporal dependencies, while incorporating external contextual factors such as weather, holidays, and traffic accidents through feature fusion. By deploying the framework on a cloud computing platform, the model achieves large-scale data processing and real-time adaptability, significantly enhancing its applicability in Intelligent Transportation Systems (ITS).

Experimental results demonstrate that the proposed GNN–Transformer hybrid model outperforms baseline methods (LSTM, GRU, TCN, GCN, pure Transformer) across multiple evaluation metrics. Specifically, the model achieves a Mean Squared Error (MSE) of 320.9, a Root Mean Squared Error (RMSE) of 17.92, a Mean Absolute Error (MAE) of 10.53, and a coefficient of determination ($R^2$) of 0.90 on the test set. These results confirm that the model effectively captures the complex spatio-temporal dependencies of traffic flow data and provides efficient and scalable forecasting performance when deployed in a cloud environment. Compared to traditional methods, the proposed model not only reduces prediction error but also exhibits superior adaptability to dynamic traffic conditions.

The findings of this research hold significant practical implications. Accurate traffic flow forecasting supports traffic signal optimization, congestion management, and route planning, thereby contributing to the advancement of smart cities and ITS applications. The cloud-based deployment ensures scalability and real-time adaptability, providing technical support for large-scale implementation in urban road networks.

Despite the important findings, this study has some limitations, such as reduced accuracy in extreme traffic conditions and limited use of unstructured data. Future research could explore integrating real-time dynamic data and optimizing edge-cloud collaboration.

Future research directions include expanding the scope of external features by integrating real-time dynamic data such as weather forecasts and traffic video streams, exploring edge-cloud collaborative architectures to reduce latency, and incorporating emerging techniques such as reinforcement learning and causal inference to improve decision-making and interpretability under complex conditions. With these improvements, the proposed cloud-based GNN–Transformer hybrid framework has the potential to achieve broader and deeper applications in intelligent transportation.

In conclusion, this study, through a GNN–Transformer hybrid approach with external feature fusion and cloud deployment, reveals a scalable and accurate solution for traffic prediction, providing new insights for ITS development.